# Electron microscopy probing electron-photon interactions in SiC nanowires with ultra-wide energy and momentum match

Jinlong Du [1, †], Jin-hui Chen [2, 3, †], Yuehui Li [1, 4, †], Ruochen Shi [1, 4], Mei Wu [1, 4], Yun-Feng Xiao [2, *], Peng Gao [1, 4, 5, 6, *]

1. Electron Microscopy Laboratory, School of Physics, Peking University, Beijing, China.

2. State Key Laboratory for Mesoscopic Physics and Frontiers Science Center for Nano-optoelectronics, School of Physics, Peking University, Beijing 100871, China.

3. Institute of Electromagnetics and Acoustics, Xiamen University, Xiamen 361005, China

4. International Center for Quantum Materials, Peking University, Beijing, China.

5. Collaborative Innovation Center of Quantum Matter, Beijing, China.

6. Interdisciplinary Institute of Light-Element Quantum Materials and Research Center for Light-Element Advanced Materials, Peking University, Beijing 100871, China.

* Corresponding author. Email: yfxiao@pku.edu.cn (Y.F. X.); p-gao@pku.edu.cn (P. G.)

† J.L. Du, J.-h. Chen and Y.H. Li contributed equally to the work.

**Nanoscale materials usually can trap light and strongly interact with it leading to many photonic device applications. The light-matter interactions are commonly probed by optical spectroscopy, which, however, have some limitations such as diffraction-limited spatial resolution, tiny momentum transfer and non-continuous excitation/detection. In this work, using scanning transmission electron microscopy-electron energy loss spectroscopy (STEM-EELS) with ultra-wide energy and momentum match and sub-nanometer spatial resolution, we study the optical microcavity resonant spectroscopy in a single SiC nanowire. The longitudinal Fabry-Perot (FP) resonating modes and the transverse whispering-gallery modes (WGMs) are simultaneously excited and detected, which span from near-infrared (~ 1.2 μm) to ultraviolet (~ 0.2 μm) spectral regime and the momentum transfer can be ranging up to $10^8$ cm$^{-1}$. The size effects on the resonant spectra of nanowires are also revealed. Moreover, the nanoscale decay length of resonant EELS is revealed, which is contributed by the strongly localized electron-photon interactions in the SiC nanowire. This work provides a new alternative technique to investigate the optical resonating spectroscopy of a single nanowire structure and to explore the light-matter interactions in dielectric nanostructures, which is also promising for modulating free electrons via photonic structures.**

**Main**

Nanophotonics, which focuses on the behavior of light at nanoscale and the interaction of light with nanomaterials, helps reduce the size of various detectors and photonic devices, as small detectors tend to have a variety of desirable properties, including low noise, high speed and low power, etc. An important direction of nanophotonics is to trap light with nanomaterials [1,2]. The trapped light enabling subwavelength optics beyond the diffraction limit, can enhance the light-matter interactions, and thus find significant applications in sub-diffraction imaging, nanoscale lasing, sensing, and surface-enhanced spectroscopy[1-4]. Understanding the light-matter interactions at the nanoscale is always prerequisite for such photonic device design and applications. The laser-based optical spectroscopies such as the absorption/scattering spectroscopy, photoluminescence spectroscopy are proved to be sophisticated tools for probing photonic response of nanomaterials and nanostructures. For example, the optical near-field techniques are widely used to image the evanescent field of the nanostructures in the most advanced set-ups[5,6]. However, there are also a few limitations for the conventional optical spectroscopy. Firstly, due to the tiny momentum for photons, only those interactions with small momentum transfer can be excited[7-12]. Second, some frequency windows (e.g., far infrared) are not covered by the current commercial laser instruments[13-16]. Moreover, optical near-field techniques need to launch continuous wave laser light spanning ultra-wide frequency spectra to excite different-order waveguide modes, making not only the experimental study technically challenging, but also the study of different waveguide modes coupling difficult[13-16]. Third, the spatial resolution of optical spectroscopy is constrained by the classical diffraction effects[17-19]. The interaction cross-sections between light and matter is typically small, thus it is usually difficult to probe the signal from the tiny nanostructures due to the limited spatial resolution and/or small cross-sections[1,20,21]. Therefore, it is highly desirable to overcome these limitations and to probe the fundamental interactions in tiny photonic nanostructures more efficiently.

Compared to the photons in optical techniques, the transferred momentum of electron can be much larger when the injected electrons interact with matter[8,22,23]. The transferred momentum of a high energy electron beam scattered at angle ($\theta$) is several orders of magnitude larger than that of a laser beam, which can be calculated by $\hbar\Delta k = \hbar k \sin\theta \cong \hbar k\theta$. For example, an electron beam with collection semi-angle of 25 mrad with wavelength $\lambda_e$=0.048 Å (60 keV) can give a transferred momentum $\Delta k$ up to $3.0\times10^8$ cm$^{-1}$. In contrast, a laser beam with a divergence of 25 mrad at wavelength 600 nm gives a $\Delta k$ of about $2.6\times10^3$ cm$^{-1}$. In addition, the range of transferred energy $\Delta E$ of the electrons is up to some kilo-electronvolt, which is much larger than that in laser beam as well.

The electron energy loss spectroscopy (EELS) incorporated in a scanning transmission electron microscope (i.e., STEM-EELS) equipped with aberration corrector and monochromator offers considerable energy resolution and very high spatial resolution. The recent advances of monochromator in STEM enable an atom-wide kilo-electronvolt electron probe with about sub-10 meV energy resolution[24-30], allowing atomically resolved EELS analysis of many physical excitations such as vibrational spectroscopy, phonon polaritons and plasmons in an extremely wide continuous spectral range with subnanometer spatial resolution[1,13,31-37]. Besides, the high energy electrons can usually excite more multipole modes even including the optical inactive dark modes[9,38]. However, the research on the optical resonating spectroscopy of dielectric nanostructures via STEM-EELS has been rarely reported, which is mainly due to the optical frequency windows of the resonating modes are usually overlapped with plasmon/interband transitions of dielectric materials in EELS. In fact, for the dielectric microcavities with suitable dimension (few tens of nanometers to micrometers), the intrinsically optical resonating spectra can be obtained without influence of plasmon and interband transitions via aloof geometries (the electron beam positioned in the vacuum near the sample)[27,39].

In this work, we characterize the interactions of fast electrons and optical resonating modes in SiC nanowires with different diameters using STEM-EELS in aloof mode where the electron beam is positioned several nanometers away from the sample. The longitudinal Fabry–Perot (FP) resonating modes and the transverse whispering-gallery modes (WGMs) of SiC nanowires are simultaneously excited and detected, which span from near-infrared (~ 1.2 μm) to ultraviolet (~ 0.2 μm) and is much wider than that of conventional optical spectroscopy. The detected momentum transfer can be up to $10^8$ cm$^{-1}$, which is also a few orders of magnitude larger than that of conventional optical spectroscopy. We systematically study many nanostructures to extract the size effect and reveal the underlying mechanism from both the analytical and numerical simulations. Moreover, the nanoscale decay length of resonant EELS is demonstrated benefited from the sub-nanometer spatial resolution of STEM-EELS, which is contributed by the strongly localized electron-photon interactions in the SiC nanowire. This work provides potential to investigate the resonant spectral properties at nanoscale via STEM-EELS in ultrawide energy and momentum range.

**Results**

Figure 1a shows a schematic of STEM-EELS operation. The 60 kV focused electron beam with 20 mrad convergence semi-angle is positioned ~ 10 nm away from the SiC nanowire, and the EELS data is collected with 25 mrad collection semi-angle. Figure 1b shows a low-magnification high angle annular dark field STEM (HAADF-STEM) micrograph of a single SiC nanowire. The diameter of the nanowire is ~246 nm with finite length (length to diameter ratio is ~9.8). No porosities or dislocations are observed in the nanowire. The electron diffraction pattern along the [110] direction of the nanowire reveals the single crystal structure without any grain boundaries (See Fig. S1a). Figure 1c shows an atomic-resolution HAADF-STEM image of the nanowire. Based on electron diffraction pattern and the atomic image, the crystal structure of the SiC nanowire is confirmed to be 3C-SiC (Fig. S1b). The two-dimensional space-EELS map of a SiC nanowire is shown in Fig. 1d, which is acquired with the electron beam located at different positions in the aloof mode by scanning from s1 to s2 along the nanowire. The spectral intensity distribution along this line clearly shows interference fringe. Unfolding the EELS spectra at a fixed position near the SiC nanowire, two groups of resonating peaks, i.e. peaks A and B are resolved as shown in Fig. 1e. The peaks that have significantly stronger intensity and larger peak energy intervals are regarded as peaks B, while the dense fringes with weaker modulated intensity and smaller peak energy intervals, are peaks A. The peaks A are stacked with peaks B and can be more clearly observed at the peaks B's valley regions, which are highlighted by red lines and one is enlarged as shown in Fig. 1f. Besides, no obvious intensity decreases of peaks A from the SiC nanowire tips to middle positions is observed, which is also verified in another nanowire with larger length to diameter ratio of ~22.5 (see Fig. S2). The energies of peaks B does not change with beam shift along the SiC nanowire. The energy intervals of peaks B decrease with the increase of energy and more than nine peaks B can be distinguished in the energy of 0-6 eV in this SiC nanowire.

To decipher the origin of peaks A, we convert the spatial field modulation of peaks A to the momentum space, i.e., $k = \pi/d_{\text{EELS}}$, where $d_{\text{EELS}}$ is distance of the spatial interference fringe as shown in inset of Fig. 2a. Without loss of generality, the dispersion relationships in SiC nanowire with different diameters can be extracted from the experimental two-dimensional EELS map, as shown in Fig. 2a. The dispersion curves show that the resonating energy increases as the increase of $k$. From the numerical calculations using COMSOL Multiphysics, it is found that the peaks A of the EELS can be attributed to the longitudinal waveguide modes, which forms FP resonance through the endface reflecting feedbacks. The dielectric dispersion of 3C-SiC is based on previous studies[40]. Detailed description of the analytical model can be seen in Methods. Figure 2b shows the schematic of the excitation and measurement of FP resonating modes in SiC nanowires by electron beams. Free electrons pass near the SiC nanowire and interact with optical excitations. The nanowire plays as an optical cavity and increases the evanescent tails in vacuum that interact with electrons[36]. Inelastic scattering of electrons and optical excitations of SiC nanowires naturally cause changes to the free-electron state, leading to the measured axial resonance spectroscopy. The calculated dispersion relationships of four different transverse electric (TE), transverse magnetic (TM) and hybrid electric (HE) waveguide modes, in a SiC nanowire with diameter of 246 nm are plotted in Fig. 2c, i.e., $TE_{01}$, $TM_{01}$, $HE_{21}$ and $HE_{12}$, in which the subscripts are used to complete the description of the waveguide modes. The typical field interference pattern of the $TE_{01}$, $TM_{01}$ and $HE_{12}$ modes can be found in Fig 2d. The calculated dispersion relationships agree reasonably well with the experimental results, confirming that the peak A is resulted from the longitudinal propagating of different waveguide modes with FP resonating effects. Note that the size effects on the excitation of resonant waveguide modes are also investigated and it shows that the energy of the first detected dispersion line decrease with the increase of nanowires diameter (Fig. 2a), which agrees with the evanescent field properties of waveguide modes in nanowires.

The features and origin of peak B are discussed in Fig. 3. Figure 3a shows the typical background-subtracted EELS profiles of SiC nanowires with different diameters. Since the energy of peak B does not depend on the axial position of the SiC nanowires, we superimposed the EELS profiles in the axial direction and normalized the intensity to reduce the errors of the peaks positions. It shows that the energy intervals of the peaks decrease with the increase of nanowires diameter, and the measurable peaks numbers increase accordingly. Combined with the numerical calculation, we found the peaks B originate from the excitation of the transverse WGMs. Figure 3b shows the schematic of free electrons interacting with nanowires by the excitation of WGM. The SiC nanowire plays as a cylindrical cavity and the WGMs with different orders can be efficiently excited due to the large momentum and energy transfer in the interaction of fast electrons and photonic resonant fields. Due to the high quality factor nature of WGMs compared with the FP modes, the spectral intensity of peaks B are significantly larger than that of peaks A. The measured resonating EELS energy can be corresponding to different orders of WGMs. Figure 3c shows the dependence of the experimental (solid dots) and calculated (hollow) energies of peak B to the WGMs on SiC nanowires of different diameters. The simulated results agree well with experimental measurements. The features of these eigen-modes with different orders, i.e., $m$ = 0, 1, 2, 3, are plotted by the cross-sectional field distribution, as shown in Fig. 3d. The simultaneous detection of these high order cavity modes essentially originates from the wide momentum and energy match in electron-mater interactions.

Compared with optical methods, STEM can easily achieve ultrahigh spatial resolution, so we are able to probe the spatial distribution of different WGMs. A SiC nanowire with diameter of ~739 nm (Fig. 4a) is selected and STEM-EELS measurements are performed near its surface, as shown in the green box in Fig. 4a. The EELS signal of each spatial location and the corresponding HAADF image are recorded at the same time during scanning in real space. The enlarged STEM image in Fig. 4b and the corresponding EELS with different peaks (Fig. 4c) near the surface of SiC nanowires show the spatial distribution of peaks B. We find that the intensities of peaks B decrease with the increase of distance to SiC nanowire, and the higher energy modes decrease more quickly. Quantitatively, Fig. 4d plots the EELS profiles along the radial direction of a SiC nanowire for different WGMs. The EELS intensity profiles of these WGMs are in accordance with exponential function $I_m = \exp(-d/\lambda_m)$, where $I_m$ is the normalized intensity of EELS peaks B at distance $d$, and $\lambda_m$ is the decay length describing the distance from the edge to the position where the field decreases by a factor of $1/e$. The decay length of these EELS peaks B are plotted in Fig. 4e. It shows that the decay length degreases with the increase of energy, i.e., the higher order WGMs, the smaller decay length of EELS profiles. Theoretically, the EELS profiles are determined by the correlation of WGMs field and the electron dipole radiation[41] (see Methods). Our numerical calculation is well consistent with the experimental results.

In summary, the electron-photon interactions in individual SiC nanowires are investigated via STEM-EELS, and two types of optical cavity resonating modes, supported by the longitudinal FP resonating waveguide modes and the transverse WGMs, are simultaneously excited through the inelastic electrons scattering, which spans from ultraviolet to near infrared spectra. The electron-photon interactions exhibit much wider energy and momentum match than photon-photon, thus the single nanowire cavity modes excited by electrons may exhibit potential applications in the design of nanoscale optoelectronic devices. Our work provides a new alternative technique to study the optical resonating spectroscopy of a single nanowire and to explore the light-matter interactions in dielectric nanostructures, which is also promising for modulating free electrons via photonic structures.

**Acknowledgements**

This work was supported by the National Natural Science Foundation of China (Grant No. 52125307, 11974023, 12004010, U20B2025, 62005231). The authors acknowledge Electron Microscopy Laboratory of Peking University, China for the use of Cs corrected Nion U-HERMES200 scanning transmission electron microscope, and financial support. We thank Dr. Chenglong Shi and Dr. Tracy Lovejoy for assistance in microscope operation.

 **Author contributions**

P. G. and Y. X. conceived the research. J. D., J. C. and Y. L. performed the main experiments and data analysis. M. W. performed experiments. J. D., J. C., Y. L., P. G. and Y. X. wrote the manuscript with inputs from all authors.

**Competing interests:** None declared. Data and materials availability: All data needed to evaluate the conclusions in the paper are present in the paper or the Supplementary Information.

## Data availability

The data that support the findings of this study are available from the corresponding author upon request.

**Methods**

**EELS data acquisition.** The EELS data was acquired on a Nion U-HERMES200 microscope equipped with both a monochromator and aberration correctors, with 60 kV work voltage and ~20 pA beam current. The beam convergence semi-angle of 20 mrad was used while the collection semi-angle is 25 mrad. The spot size of electron beam is about 200 pm, and scanned with the line positioned ~ 10 nm away from the SiC nanowire edge.

**Theoretical modeling of optical modes in SiC nanowire.** Let us first consider the longitudinal propagating modes in an infinite cylindrical SiC nanowire with radius of $a$. The electromagnetic fields in the nanowire are expressed in cylindrical coordinates as $\vec{E}=\vec{E}(\vec{r})e^{i(\omega t-\beta z)}$, $\vec{H}=\vec{H}(\vec{r})e^{i(\omega t-\beta z)}$ where $\omega$ is the angular frequency, $\beta$ is the propagation constant. We can obtain two sets of wave equations derived from the Maxwell equations:

$$\begin{cases} \dfrac{\partial^2 E_z}{\partial r^2}+\dfrac{1}{r}\dfrac{\partial E_z}{\partial r}+\dfrac{1}{r^2}\dfrac{\partial^2 E_z}{\partial \theta^2}+[k_0^2\varepsilon(\vec{r})-\beta^2]E_z=0 \\ \dfrac{\partial^2 H_z}{\partial r^2}+\dfrac{1}{r}\dfrac{\partial H_z}{\partial r}+\dfrac{1}{r^2}\dfrac{\partial^2 H_z}{\partial \theta^2}+[k_0^2\varepsilon(\vec{r})-\beta^2]H_z=0 \end{cases} \quad (1),$$

where $k_0$ is the vacuum wave-vector, $\varepsilon(\vec{r})$ is the relative permittivity of the SiC nanowire and its surroundings, i.e. $\varepsilon(\vec{r})=\varepsilon_1=\varepsilon_{SiC}(\omega)$, $r\leq a$; $\varepsilon(\vec{r})=\varepsilon_2=1$, $r>a$. Given the boundary condition that the tangential electric/magnetic field components should be continuous at the interface, and the fact that field is convergent at $r=0$ and $r=\infty$, we can obtain the eigen-equation for different waveguide modes[34,42]:

$$\left[\dfrac{J_n'(u)}{uJ_n(u)}+\dfrac{K_n'(w)}{wK_n(w)}\right]\left[\dfrac{J_n'(u)}{uJ_n(u)}+\dfrac{\varepsilon_2}{\varepsilon_1}\dfrac{K_n'(w)}{wK_n(w)}\right]=n^2\left(\dfrac{1}{u^2}+\dfrac{1}{w^2}\right)\left(\dfrac{1}{u^2}+\dfrac{\varepsilon_2}{\varepsilon_1}\dfrac{1}{w^2}\right) \quad (2)$$

where $n$ is integer (0, 1, 2, 3…), $J_n(x)$ is the $n$th-order Bessel function, $K_n(x)$ is modified Bessel functions of the second kind, $u$ and $w$ are defined as:

$$u=a\sqrt{\varepsilon_1 k_0^2-\beta^2},\ w=a\sqrt{\beta^2-\varepsilon_2 k_0^2} \quad (3).$$

When $n=0$, equation (2) is reduced to the following equations:

$$\begin{cases} \dfrac{J_0'(u)}{uJ_0(u)} + \dfrac{K_0'(w)}{wK_0(w)} = 0, & \text{TE} \\[2ex] \dfrac{J_0'(u)}{uJ_0(u)} + \dfrac{\varepsilon_2}{\varepsilon_1}\dfrac{K_0'(w)}{wK_0(w)} = 0, & \text{TM} \end{cases} \qquad (4).$$

The solutions to equation (4) generate the transverse electric TE (transverse magnetic TM) mode family, where the longitudinal electric field $E_z$ (magnetic field $H_z$) is zero. When $n \neq 0$, the solution to equation (2) generate the HE and EH modes. The HE (EH) modes family indicate the longitudinal electric (magnetic) field is relatively strong.

In equation (1), when $\beta=0$, it indicates that the wave propagates totally in the transverse plane, which is also considered as WGM, and we have the following equations:

$$\begin{cases} \dfrac{\partial^2 E_z}{\partial r^2} + \dfrac{1}{r}\dfrac{\partial E_z}{\partial r} + \dfrac{1}{r^2}\dfrac{\partial^2 E_z}{\partial \theta^2} + k_0^2 \varepsilon(\vec{r}) E_z = 0 \\[2ex] \dfrac{\partial^2 H_z}{\partial r^2} + \dfrac{1}{r}\dfrac{\partial H_z}{\partial r} + \dfrac{1}{r^2}\dfrac{\partial^2 H_z}{\partial \theta^2} + k_0^2 \varepsilon(\vec{r}) H_z = 0 \end{cases} \qquad (5).$$

The electromagnetic fields can be expressed as $\vec{E} = \vec{E}(\vec{r})e^{-im\theta}, \vec{H} = \vec{H}(\vec{r})e^{-im\theta}$, $m$ is an integer. Similarly, we can obtain the following eigen-equation by considering the boundary conditions [43]:

$$\left[\sqrt{\varepsilon_1}\dfrac{J_m(\sqrt{\varepsilon_1}k_0 a)}{J_m'(\sqrt{\varepsilon_1}k_0 a)} - \sqrt{\varepsilon_2}\dfrac{H_m^{(1)}(\sqrt{\varepsilon_2}k_0 a)}{H_m^{'(1)}(\sqrt{\varepsilon_2}k_0 a)}\right]\left[\sqrt{\varepsilon_2}\dfrac{J_m(\sqrt{\varepsilon_1}k_0 a)}{J_m'(\sqrt{\varepsilon_1}k_0 a)} - \sqrt{\varepsilon_1}\dfrac{H_m^{(1)}(\sqrt{\varepsilon_2}k_0 a)}{H_m^{'(1)}(\sqrt{\varepsilon_2}k_0 a)}\right] = 0 \qquad (6),$$

where $H_m^{(1)}(x)$ is the Hankel function of the first kind. The solutions to equation (6) can be further divided into two groups, i.e. whispering-gallery TE (WTE) and whispering-gallery TM (WTM):

$$\begin{cases} \sqrt{\varepsilon_1}\dfrac{J_m'(\sqrt{\varepsilon_1}k_0 a)}{J_m(\sqrt{\varepsilon_1}k_0 a)} - \sqrt{\varepsilon_2}\dfrac{H_m^{'(1)}(\sqrt{\varepsilon_2}k_0 a)}{H_m^{(1)}(\sqrt{\varepsilon_2}k_0 a)} = 0, & \text{WTE} \\[2ex] \sqrt{\varepsilon_2}\dfrac{J_m'(\sqrt{\varepsilon_1}k_0 a)}{J_m(\sqrt{\varepsilon_1}k_0 a)} - \sqrt{\varepsilon_1}\dfrac{H_m^{'(1)}(\sqrt{\varepsilon_2}k_0 a)}{H_m^{(1)}(\sqrt{\varepsilon_2}k_0 a)} = 0, & \text{WTM} \end{cases} \qquad (7).$$

Note that the WTE (WTM) polarization modes indicate that the electric field component (magnetic field component) is only $E_z$ ($H_z$).

The interaction of electron beams and SiC nanowires can be treated as follows: the dipole radiation of a moving electron induced the polarization field of a photonic nanowire, which in turn acts on the electron and brings the energy loss. The generated electron energy loss can be expressed as [34,41]:

$$\Gamma_{EELS}(\omega) = \frac{e^2}{\pi\hbar\omega} \iint cos\left[\frac{\omega(z-z')}{v}\right] \times Im\left[\frac{E_z^{ind}(r,z,\omega)}{p_z(r',z',\omega)}\right] dz dz' \qquad (8).$$

Here we suppose the electron with velocity of $v$ moves along $z$ direction, and the electron is modeled as an electric dipole of amplitude $p_z(z,\omega)$. $E_z^{ind}(z,\omega)$ is the electron-induced electric field of the nanowire structure; $-e$ is the electron charge, $\hbar$ is the reduced Planck's constant, $\omega$ is the angular frequency. Under the mean field approximation, the Equation (8) can be rewritten as:

$$\Gamma_{EELS}(\omega) \approx \frac{e^2}{\pi\hbar\omega} A \times Im\left[\frac{E_z^{ind}(r_0,z_0,\omega)}{p_z(r_0,z_0,\omega)}\right] \qquad (9)$$

where A is the effective interaction area, $r_0$ is the average interaction distance. We expand the $E_z^{ind}(r_0,z_0,\omega)$ by the eigen-photonic-mode of the nanowires and obtained that $E_z^{ind}(r_0,z_0,\omega) = \alpha(\omega)E_z^{eigen}(r_0,z_0,\omega)$, where $\alpha(\omega)$ is the excited field amplitude and is calculated through the spatial overlap integral of dipole field and a photonic-eigen mode:

$$\alpha(\omega) = \iint p(\vec{r},z_0,\omega) \cdot E_z^{eigen,*}(\vec{r},\omega) d\vec{r} \qquad (10).$$

Note that $E_z^{eigen}(r_0,z_0,\omega)$ of a cylinder nanowire structure has analytical expressions as explicitly discussed in the first section. Finally, substituting these expressions into Equation (9), we obtained the electron loss spectra as:

$$\Gamma_{EELS}(\omega) = \frac{e^2}{\pi\hbar\omega} A \times Im\left[\frac{\alpha(\omega)E_z^{eigen}(r_0,z_0,\omega)}{p_z(r_0,z_0,\omega)}\right] \qquad (11)$$

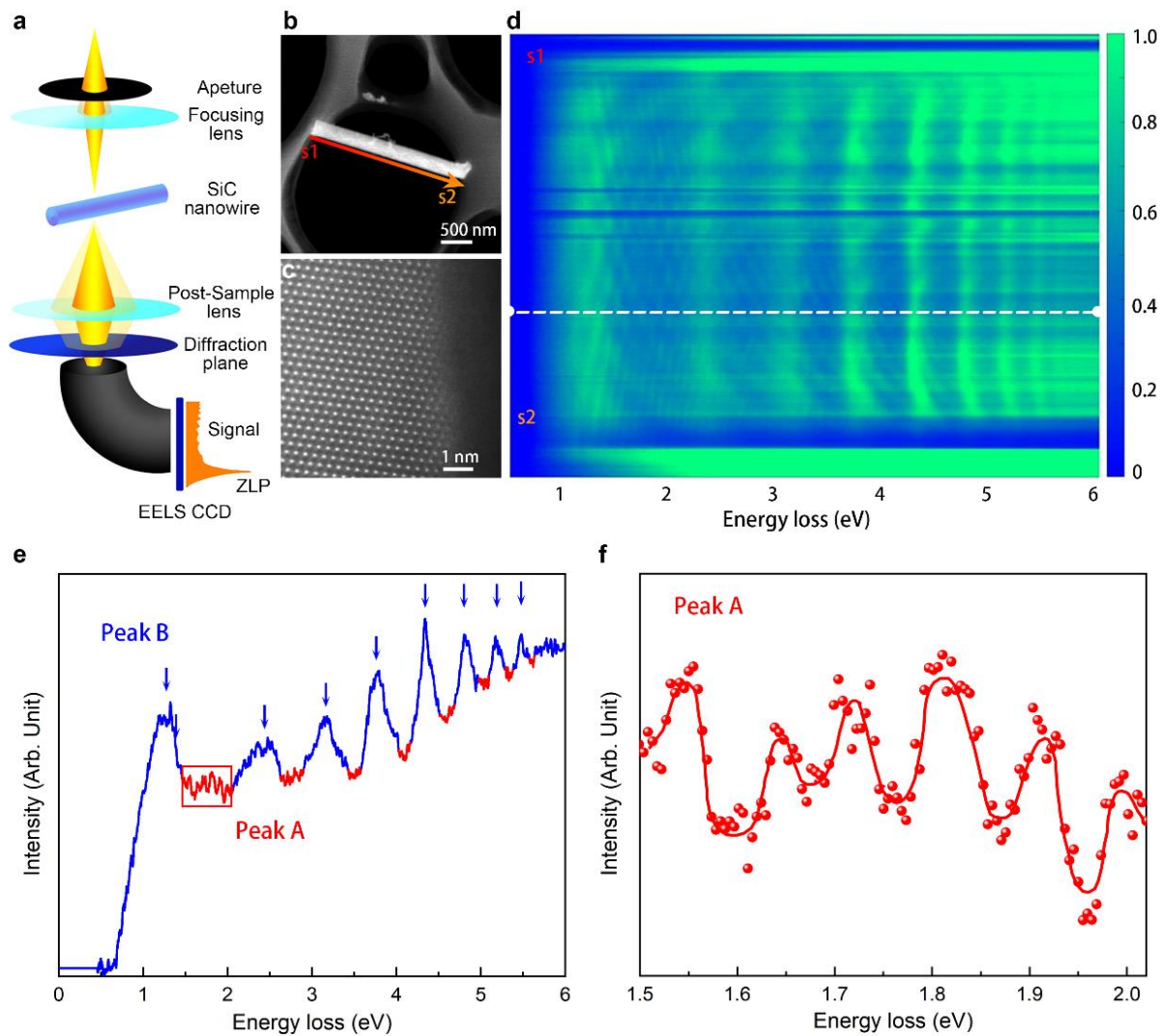

**Fig. 1 STEM-EELS measurements of 3C-SiC nanowires. a** Schematic diagram of STEM-EELS system. **b** HAADF-STEM micrograph of a SiC nanowire with diameter of ~246 nm and length to diameter ratio of ~ 9.8. **c** An atomic resolution HAADF-STEM image of edge of SiC nanowire viewed along [110] direction. **d** Experimental EELS mapping of the nanowire, taken by aloof electron beam scanning along nanowire from s1 to s2. **e** The EELS intensity line profile showing two types of peaks. **f** Enlarged view of the red box in **e**, showing the profile of peak A.

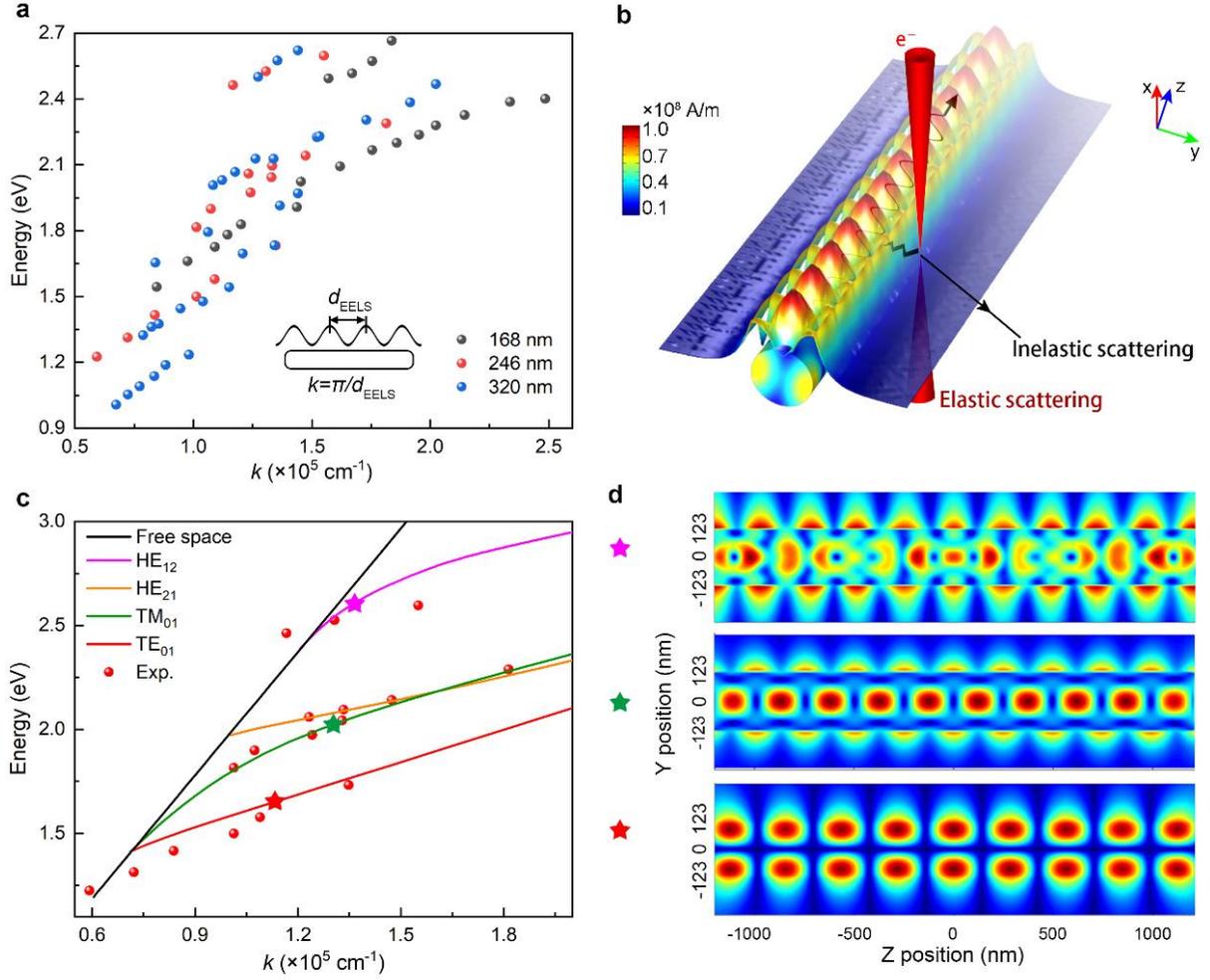

**Fig. 2 Size effects on the excitation of resonant waveguide modes of SiC nanowires. a** The dispersion of peak A in SiC nanowire with different diameters. **b** Schematic diagram showing the excitation and measurement of FP resonating modes in SiC nanowires. **c** The measured (red dot) and calculated (line) dispersion of peak A in SiC nanowire with diameter of ~246 nm. The peaks A is induced by different photonic modes, i.e. $TE_{01}$, $TM_{01}$, $HE_{21}$ and $HE_{12}$. **d** The calculated longitudinal propagating of $TE_{01}$, $TM_{01}$, and $HE_{12}$ modes with ($k=1.12\times10^5$ cm$^{-1}$, $E=1.55$ eV), ($k=1.27\times10^5$ cm$^{-1}$, $E=1.99$ eV) and ($k=1.30\times10^5$ cm$^{-1}$, $E=2.65$ eV) respectively, which are labeled by the stars with different colors in **c**.

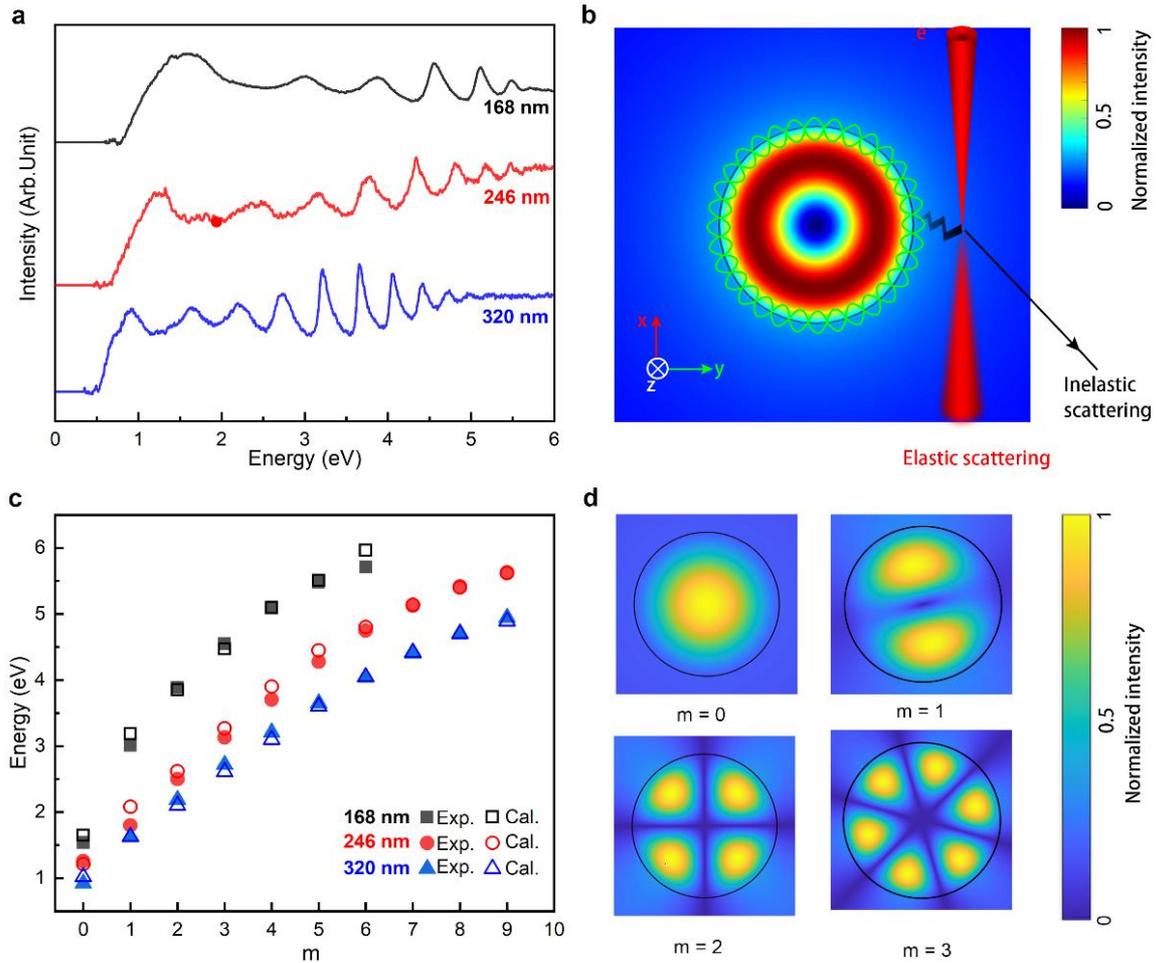

**Fig. 3 Size effects on the excitation of transverse whispering gallery mode (WGM) of SiC nanowire. a** Typical background-subtracted EELS profiles of SiC nanowires with different diameters. **b** Schematic showing interaction of free electrons with the excitation of WGM. **c** Experimental (solid dots) and calculated (hollow dots) energies of peaks B on SiC nanowires of different diameters. **d** The simulated two-dimensional cross-sectional field distribution of first four whispering-gallery modes (mode order $m$=0-3) with transverse magnetic polarization in a nanowire with diameter of 320 nm.

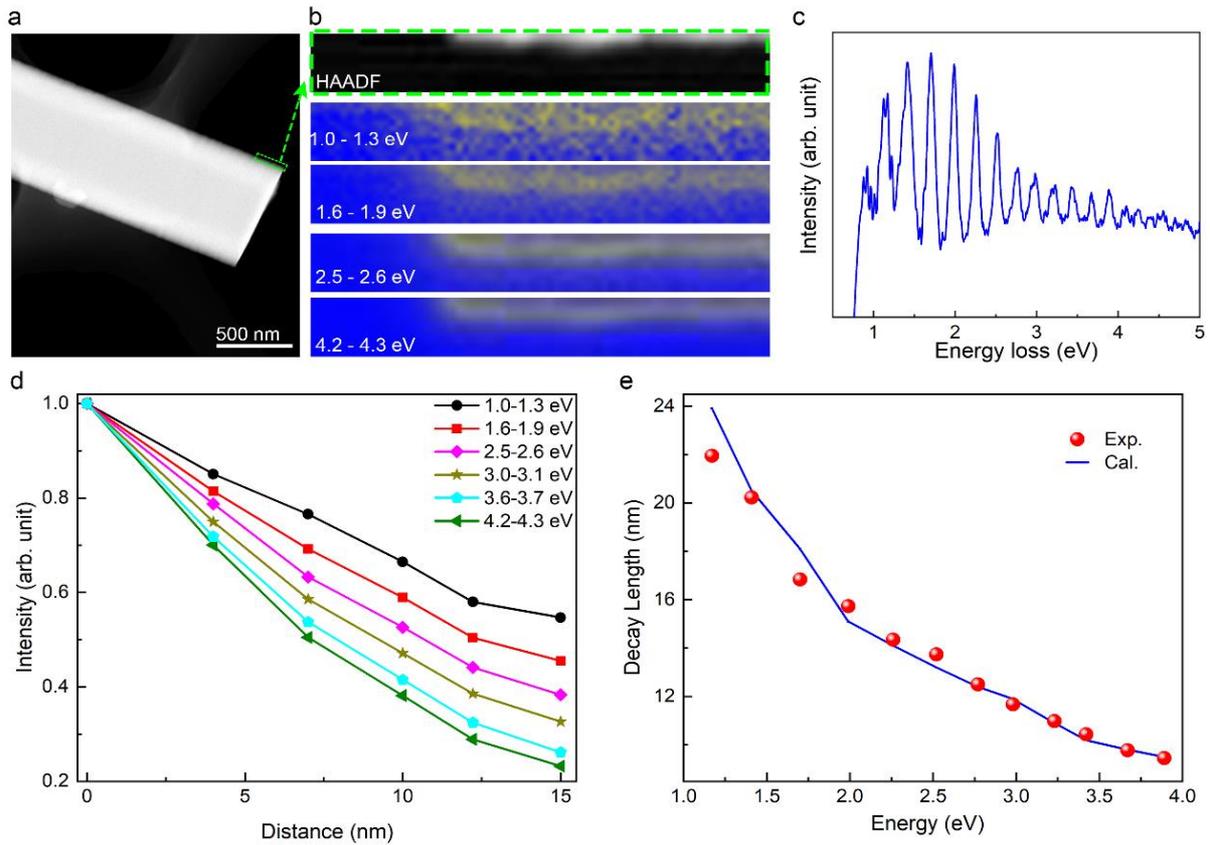

**Fig. 4 Spatial characterizations of EELS spectra for peaks B. a** HAADF-STEM image of SiC nanowire with diameter of ~739 nm. **b** Experimental EELS maps near the surface of a SiC nanowire showing the spatial distribution of peaks B of various energy. **c** The background-subtracted EELS profiles of the SiC nanowire, showing the peaks supported from different orders of WGMs. **d** Normalized intensity profile showing the decay features of peaks B along the radial direction of SiC nanowires. **e** The experimental and calculated decay length of peaks B.

# Supplementary information for:

# Electron microscopy probing electron-photon interactions in SiC nanowires with ultra-wide energy and momentum match

Jinlong Du [1, †], Jin-hui Chen [2, 3, †], Yuehui Li [1, 4, †], Ruochen Shi [1, 4], Mei Wu [1, 4], Yun-Feng Xiao [2, *], Peng Gao [1, 4, 5, 6, *]

*1. Electron Microscopy Laboratory, School of Physics, Peking University, Beijing, China.*

*2. State Key Laboratory for Mesoscopic Physics and Frontiers Science Center for Nano-optoelectronics, School of Physics, Peking University, Beijing 100871, China.*

*3. Institute of Electromagnetics and Acoustics, Xiamen University, Xiamen 361005, China*

*4. International Center for Quantum Materials, Peking University, Beijing, China.*

*5. Collaborative Innovation Center of Quantum Matter, Beijing, China.*

*6. Interdisciplinary Institute of Light-Element Quantum Materials and Research Center for Light-Element Advanced Materials, Peking University, Beijing 100871, China.*

* Corresponding author. Email: yfxiao@pku.edu.cn (Y.F. X.); p-gao@pku.edu.cn (P. G.)

† J.L. Du, J.-h. Chen and Y.H. Li contributed equally to the work.

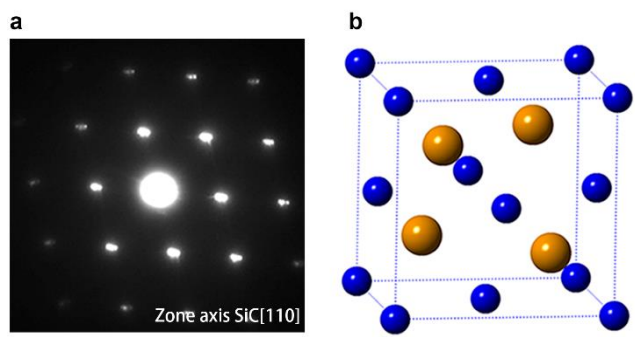

**Fig. S1. SiC selected area electron diffraction and atomic structures. a** The electron diffraction pattern of SiC along zone axis [110]. **b** Atomic structure of 3C-SiC with blue spheres for C atoms and brown spheres for Si atoms.

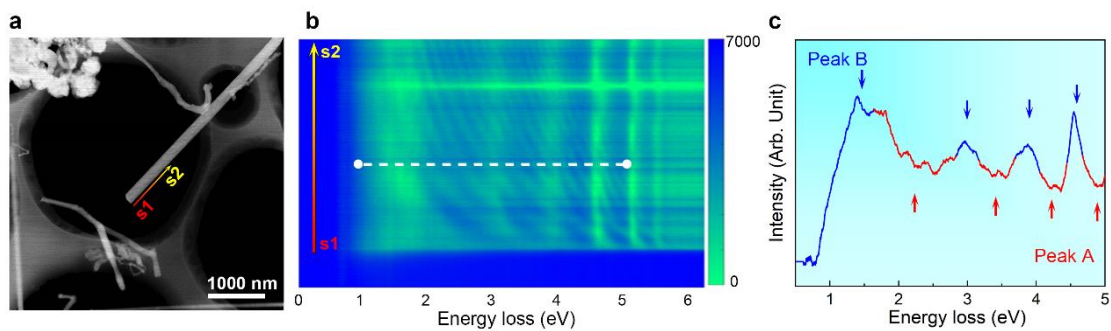

**Fig. S2 STEM imaging and EELS measurements near the terminal of 3C-SiC nanowire in aloof mode. a** HAADF-STEM image of the of 3C-SiC nanowire with diameter of 180 nm with length/diameter ratio of ~22.5. **b** Experimental two-dimensional EELS maps near the nanowire, taken by aloof beam (scanning along nanowire from s1 to s2). **c** EELS profile along the dashed line in **b**, showing two types of peaks.

Figure S2 show the STEM imaging and EELS measurements in another SiC nanowire with diameter of 180 nm and length/diameter ratio as high as ~ 22.5. Fig. S2a shows the HAADF-STEM micrographs of the selected SiC nanowire. Fig. S2b shows the measured two-dimensional space-EELS map near the nanowire, acquired with the electron beam located at different position by scanning along s1 to s2 of the nanowire. Both peaks A and peaks B are observed. The energy intervals of peaks B decrease with the increase of energy. The peaks A cannot be observed in the SiC nanowires with infinite length, indicating peak A generated from the reflection of tips of SiC nanowires. The peaks A are demonstrated to be the FP resonating modes.